\documentclass[prl,twocolumn,showpacs,amsmath,amssymb,superscriptaddress]{revtex4}

\usepackage{graphicx}
\usepackage{dcolumn}
\usepackage{bm}
\usepackage{amsmath}
\usepackage{amssymb}
\usepackage{multirow}
\usepackage{enumerate}

\begin{document}

\title{Multiplexed communication over a high-speed quantum channel}

\author{M. Heurs}
\affiliation{Centre for Quantum Computer Technology, School of Information
Technology and Electrical Engineering, University College, The
University of New South Wales, Canberra, ACT, 2600, Australia}
\author{J. G. Webb}
\affiliation{Centre for Quantum Computer Technology, School of Information
Technology and Electrical Engineering, University College, The
University of New South Wales, Canberra, ACT, 2600, Australia}
\author{A. E. Dunlop}
\affiliation{Centre for Quantum Computer Technology, School of Information
Technology and Electrical Engineering, University College, The
University of New South Wales, Canberra, ACT, 2600, Australia}
\author{C. C. Harb}
\affiliation{Centre for Quantum Computer Technology, School of Information
Technology and Electrical Engineering, University College, The
University of New South Wales, Canberra, ACT, 2600, Australia}
\author{T. C. Ralph}
\affiliation{Centre for Quantum Computer Technology, Department of Physics, The University of Queensland, St Lucia, Brisbane, QLD, 4072, Australia}
\author{E. H. Huntington}
\affiliation{Centre for Quantum Computer Technology, School of Information
Technology and Electrical Engineering, University College, The
University of New South Wales, Canberra, ACT, 2600, Australia}

\date{\today}

\begin{abstract}

In quantum information systems it is of particular interest to consider the best way in which to use the non-classical resources consumed by that system.   Quantum communication protocols are integral to quantum information systems and are amongst the most promising near-term applications of quantum information science.  Here we show that a multiplexed, digital quantum communications system supported by comb of vacuum squeezing has a greater channel capacity per photon than a source of broadband squeezing with the same analogue bandwidth.  We report on the time-resolved, simultaneous observation of the first dozen teeth in a 2.4~GHz comb of vacuum squeezing produced by a sub-threshold OPO, as required for such a quantum communications channel.  We also demonstrate multiplexed communication on that channel.

\end{abstract}


\maketitle

\section{Introduction}

Quantum information science lies at the nexus of quantum mechanics and information science \cite{qis}.  Quantum information systems will most likely comprise quantum information processing nodes connected by quantum communication channels \cite{qinternet} upon which quantum communication protocols, such as quantum key distribution \cite{bb84,grosshans02}, quantum dense coding \cite{bennett92,braunstein00} or quantum teleportation \cite{bennett,braunstein98} can be implemented \cite{buttler,lance,mattle,mizuno,bouwmeester,furusawa}.  A particularly important question in quantum information is how to make the best use of the quantum resources available, given the constraints of the system and its expected use (e.g. \cite{resource}).  Here we consider the issue of making the best use of quantum resources in the context of high-capacity, multiplexed quantum communications.  

One particularly useful optical non-classical state for quantum communications is the squeezed vacuum \cite{andersen,Vahlbruch}.  The squeezed vacuum exhibits reduced noise relative to a classical channel in one measurement quadrature, at the cost of increased noise in the orthogonal quadrature.  Simple passive operations can create entanglement, a key quantum resource, from squeezed vacua. With the addition of photon counting other key resource states such as single photons \cite{LVO00} and cat states \cite{gra07} can be heralded from squeezed vacua. Hence the study of quantum channels based on squeezed vacuum states can lead to a quite general understanding of the requirements of quantum communication channels. Here we produce and analyze such a channel.

In spite of its name, a squeezed vacuum actually carries photons \cite{walls}.  So the spectral properties of the squeezed vacuum must be well matched to the digital signaling scheme in order to avoid consuming non-classical resources (i.e. photons) unnecessarily.   Irrespective of the details of the coding protocol, contemporary digital communication schemes must all allow multiple users to have access to a high-capacity channel without experiencing interference from other users \cite{sklar}.  

We shall show theoretically that a comb of squeezing can support a greater multiplexed channel capacity per photon than a source of broadband squeezing with the same analogue bandwidth.  We report on the time-resolved, simultaneous observation of the first dozen teeth in a greater than 2.4~GHz comb of vacuum squeezing produced by a sub-threshold optical parametric oscillator (OPO) and we demonstrate frequency-division multiplexed (FDM) communication on that channel.  Combs of squeezing have been shown to be useful as the basic resource in creating cluster states for one-way quantum computing with continuous variables \cite{Zaidi}, but here we shall focus on their utility as a resource for quantum communication protocols.   

\section{Theory}

The Shannon capacity \cite{shannon} of a communication channel with Gaussian noise (signal) of variance $V_n$ ($V_s$) operating at the bandwidth limit is

\begin{equation}
\label{capacity}
C=\frac{1}{2}\log_2 \left[ 1+V_s/V_n \right] .
\end{equation}

Eq. \ref{capacity}  can be used to calculate the channel capacities
of quantum states with Gaussian probability distributions  \cite{ralph02,caves}.  

The quality of a quantum channel is quantified by the bandwidth-limited channel capacity for a given consumption of non-classical resources, in this case photon number.   The mean photon number per bandwidth per second of a light beam is related to the normalised variances in both the amplitude ($V^+$) and phase ($V^-$) quadratures \cite{ralph00} as:

\begin{equation}
\label{n}
\bar{n}(\omega)=\frac{1}{4} \left[ V^+(\omega) + V^-(\omega) -2\right].
\end{equation}

The variance in the encoded quadrature is  $V_{ne}=V^-+V_s$, while the variance in the unencoded quadrature is $V_{nu}=1/V^-$, assuming minimum uncertainty states and phase quadrature encoding.  

We are interested in determining the maximum channel capacity for a given mean photon flux in the physical channel.  The mean photon flux is given by $\Phi=\int \bar{n}(\omega) d\omega$ and may be combined with Eq. \ref{capacity}, Eq. \ref{n} and the power spectral densities of the squeezing and signal spectra to find the channel capacity for a given mean photon flux.  

To the signal spectrum first.  Multiple users can be given mulitplexed access to a digital communications channel via either time-division multiplexing (TDM) or FDM \cite{Lathi}.  In TDM, individual users are given access to the channel in specific time-slots, with guard intervals between each use of the channel.   In FDM, individual users are given access to specific frequency sub-bands within the channel for the duration of their access to the channel, with guard bands between each sub-band.  The guard intervals (bands) for TDM (FDM) ensure that there is no cross-talk between users.  Assuming that the channel is under more-or-less continuous use, the signal spectrum in a multiplexed, low-cross-talk, digital quantum communications system will be a continuous-wave frequency comb \cite{sklar,Lathi}.  %

First consider the usual situation of squeezing that is spectrally white over the analogue bandwidth of the quantum channel.   The signal spectrum will be a CW frequency comb, so:

\begin{eqnarray}
\Phi&=&\frac{1}{4}\int d\omega \left[V_s+(V^--1)^2/V^-\right]\nonumber\\
&=&\frac{1}{4} \left[B_{s}\tilde V_{s}+B(\tilde V^- -1)^2/\tilde V^-\right],
\end{eqnarray}

\noindent  where the normalised power spectral densities of the signal and noise are $V_s=\tilde V f(\omega)$ and $V^-=\tilde V^-$.   The function $f(\omega)$ represents the comb-like nature of the signal spectrum.   We define $B_{s}$ to be the integrated bandwidth consumed by the digital signaling scheme.  The analogue bandwidth of the quantum channel is $B \geq B_S$.  

The maximum SNR for a given $\Phi$ will be $\tilde V_s/\tilde V^-=4(\Phi^2+\Phi B)/(B B_{s})$ and the capacity for the quantum channel supported by a white squeezing spectrum is:
\begin{equation}
\label{qCapacity2}
C_{white}=\frac{1}{2}\log_2\left[(BB_{s}+4\Phi^2+4\Phi B)/(BB_{s})\right].
\end{equation}

Now consider the situation in which the squeezing spectrum is matched to the signal spectrum (i.e. the squeezing spectrum also has a comb structure) so that $V^-=\tilde V^- f(\omega)$, and thus:
\begin{eqnarray}
\label{combflux}
\Phi&=&\frac{1}{4} \left[\tilde V_{s}+(\tilde V^--1)^2/\tilde V^-\right]B_{s}.
\end{eqnarray}

For Eq. \ref{combflux}, the maximum signal-to-noise ratio (SNR) for a given $\Phi$ in the channel is $\tilde V_s/\tilde V^-=4(\Phi^2+\Phi B_{s})/B_{s}^2$, which occurs at the optimum level of squeezing $\tilde V^-_{opt}=B_{s}/(B_{s}+2 \Phi)$.  This leads to the following capacity:

\begin{equation}
\label{qCapacity}
C_{comb}=\log_2\left[1+2\Phi/B_{s}\right].
\end{equation}

The channel capacity for a comb of squeezing is always greater than that for a white squeezing spectrum when constrained to the same photon flux.  

Eq. \ref{qCapacity} is the standard result for a squeezed channel \cite{caves}.  Comparison between Eqs. \ref{qCapacity2} and \ref{qCapacity} shows that the "standard result" is in fact the optimum squeezed channel capacity, achieved only when the signal and squeezing spectra are properly matched.  

When restricted to homodyne detection, the optimum squeezed channel capacity is always greater than the capacity of coherent-state (i.e. classical) channel with the same bandwidth and photon flux.  In the notation used here, the coherent-state channel capacity with homodyne detection is $C_{coh}=\log_2(\sqrt{1+4\Phi/B_{s}})$ \cite{ralph02}.  It has been shown that communications systems using coherent states and optimal detection schemes can exceed $C_{coh}$, but as yet there is no known realization for those detection schemes (see Ref. \cite{shapiro} and the references therein).  We have restricted the analysis presented here to homodyne detection, for which there is an established experimental realization \cite{caves} and which therefore can be tested experimentally using current technology.

The ultimate capacity of a quantum communications channel exceeds the squeezed channel capacity and is set by the Holevo bound \cite{caves}.  The Holevo-bounded channel capacity can be achieved using Fock states and photon number detection \cite{caves}.  Dense coding systems constructed from squeezed vacuum sources can meet and even exceed the Holevo bound under experimentally realistic levels of squeezing and purity \cite{ralph02}.  One observation to be made from the analysis herein is that the conclusions from previous analyses of dense coding capacities still stand, but only if the signal and squeezing spectra are well matched. 
 
The output of a sub-threshold OPO has been predicted theoretically to exhibit precisely the required spectral characteristics \cite{Dunlop04}.   Frequency-resolved measurements of the first three resonances of an OPO were measured in Ref. \cite{Senior} and subsequent measurements indicated that such systems could exhibit useful squeezing at many more resonances \cite{Dunlop08}.   In neither case were multiple resonances observed and used simultaneously.  Here we present the first time-resolved, simultaneous measurements of the first dozen teeth in the comb of squeezing produced by a sub-threshold OPO, as required for a high-speed multiplexed digital quantum communications channel.

Generalising the result from Ref. \cite{Dunlop04} to include losses, the quadrature fluctuations for the channel as a function of angular frequency, $\omega$ are:

\begin{equation}
\label{outputs}
\delta X_{out}^{\mp} = \frac{ \left[ 2\kappa_{in}- \kappa \mp \chi + \left( \frac{1-e^{i\omega
\tau}}{\tau} \right) \right] \delta X_{in}^{\pm}+ 2 \sqrt{\kappa_{in}\kappa_l} \delta X_l ^{\pm}} { \kappa\pm \chi -\left(\frac{1-e^{i\omega \tau}}{\tau}\right)} ,
\end{equation}

\noindent where the cavity decay rate is $\kappa = \kappa_{in}+\kappa_l$ for $\kappa_{in}=T/\tau$ and $\kappa_l=L/\tau$.  The cavity round trip time is $\tau$ and the phase-matching bandwidth of the crystal is taken to be large compared to $1/\tau$.   The nonlinear frequency conversion rate is $\chi = 2 \beta_{in} \chi^{(2)}$ where $\chi^{(2)}$ is the second order coefficient of nonlinearity for the nonlinear material and  $\beta_{in}$ is the amplitude of the pump field (assumed to be real without loss of generality).   

Here the OPO output phase ($\delta X_{out}^-$) and amplitude ($\delta X_{out}^+$) quadrature fluctuations set the noise floor in the encoded and unencoded quadratures respectively.   The variances of the amplitude and phase quadratures, normalised to the quantum noise limit (QNL), are found from $V(\omega)=|\delta X(\omega)|^2$.    Resonances in the squeezing spectrum are separated in frequency by the cavity free-spectral range (FSR), and we can think of each squeezing resonance as contributing to the quantum channel 

\section{Experiment}

\begin{figure}[htbp]
  \centering
  \vspace{-2pt}
  \includegraphics[width=\linewidth,viewport=80 150 710 350,clip]{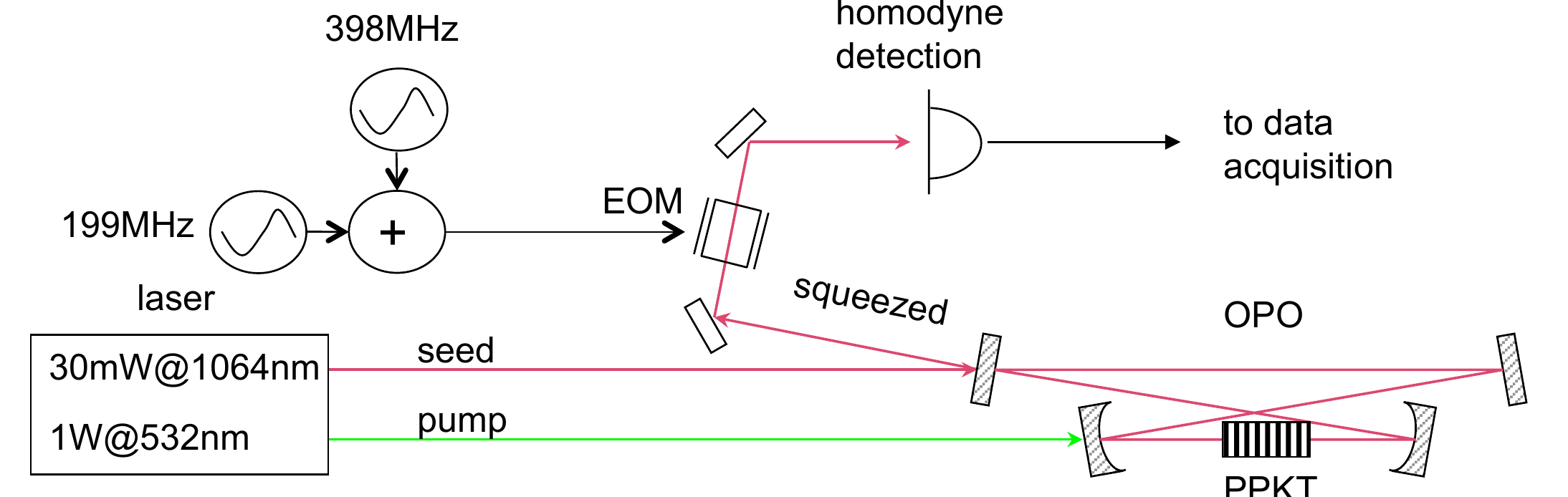}
  \caption{Schematic of experiment.}
  \label{schem}
\end{figure}

Fig. \ref{schem} illustrates schematically the experiment.  The 532nm, frequency-doubled output of a
diode-pumped, miniature monolithic Nd:YAG laser is used to pump a sub-threshold OPO.  The nonlinear crystal is periodically-poled KTP, with a
phase-matching temperature of $33.5^{\circ}$C.  The OPO has a free-spectral range of 199
MHz and is operated with a parametric amplification of 3.9 dB and de-amplification of 2.6dB.  A small fraction of the source laser power is tapped
off prior to frequency-doubling and is used as the seed to the OPO and the local oscillator (LO) for the homodyne measurements  

\begin{figure}[htbp]
  \centering
  \vspace{-2pt}
  \includegraphics[width=\linewidth, viewport=300 320 540 490,clip]{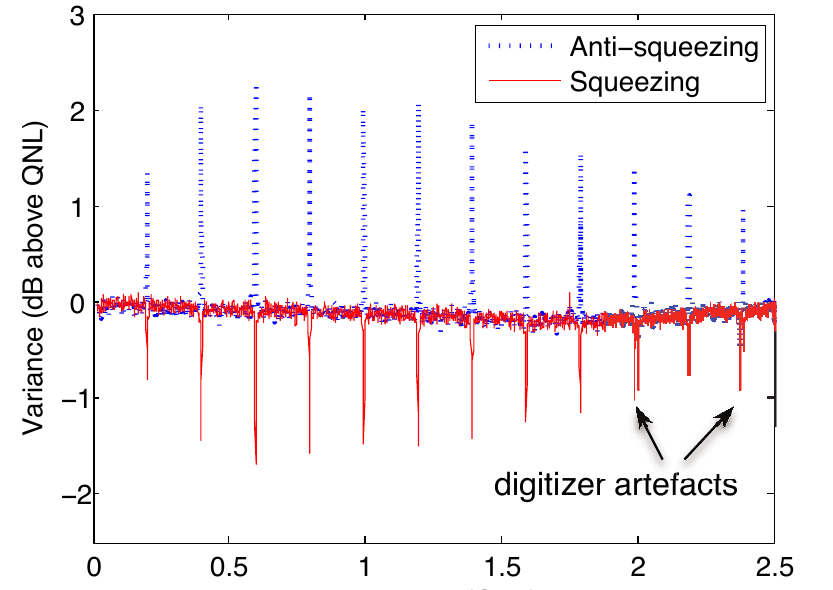}
  \caption{The averaged DFT of time-resolved quadrature homodyne measurement of squeezing comb.  Artefacts arise from the frequency response of data acquisition (color online).}
  \label{sqz}
\end{figure}

The output of the OPO is sent to a homodyne detection system with an analogue bandwidth of 2.5 GHz.   The homodyne measurement is digitally sampled at 8 GS/s, which is sufficient for time-resolved homodyne detection.  For the purposes of producing a frequency-resolved view, the discrete Fourier transforms (DFTs) of 12207 time-resolved measurements of 1.024 $\mu$s duration are computed and their magnitudes averaged.  The resulting frequency spectrum is shown in Fig. \ref{sqz}.  The variances of the squeezed and anti-squeezed output of the OPO are shown relative to the measured quantum noise limit (QNL).  The first dozen teeth in the comb of squeezing are clearly observed in Fig. \ref{sqz}.

\begin{figure}[htbp]
  \centering
  \vspace{-2pt}
  \includegraphics[width=\linewidth,,viewport=240 150 600 430,clip]{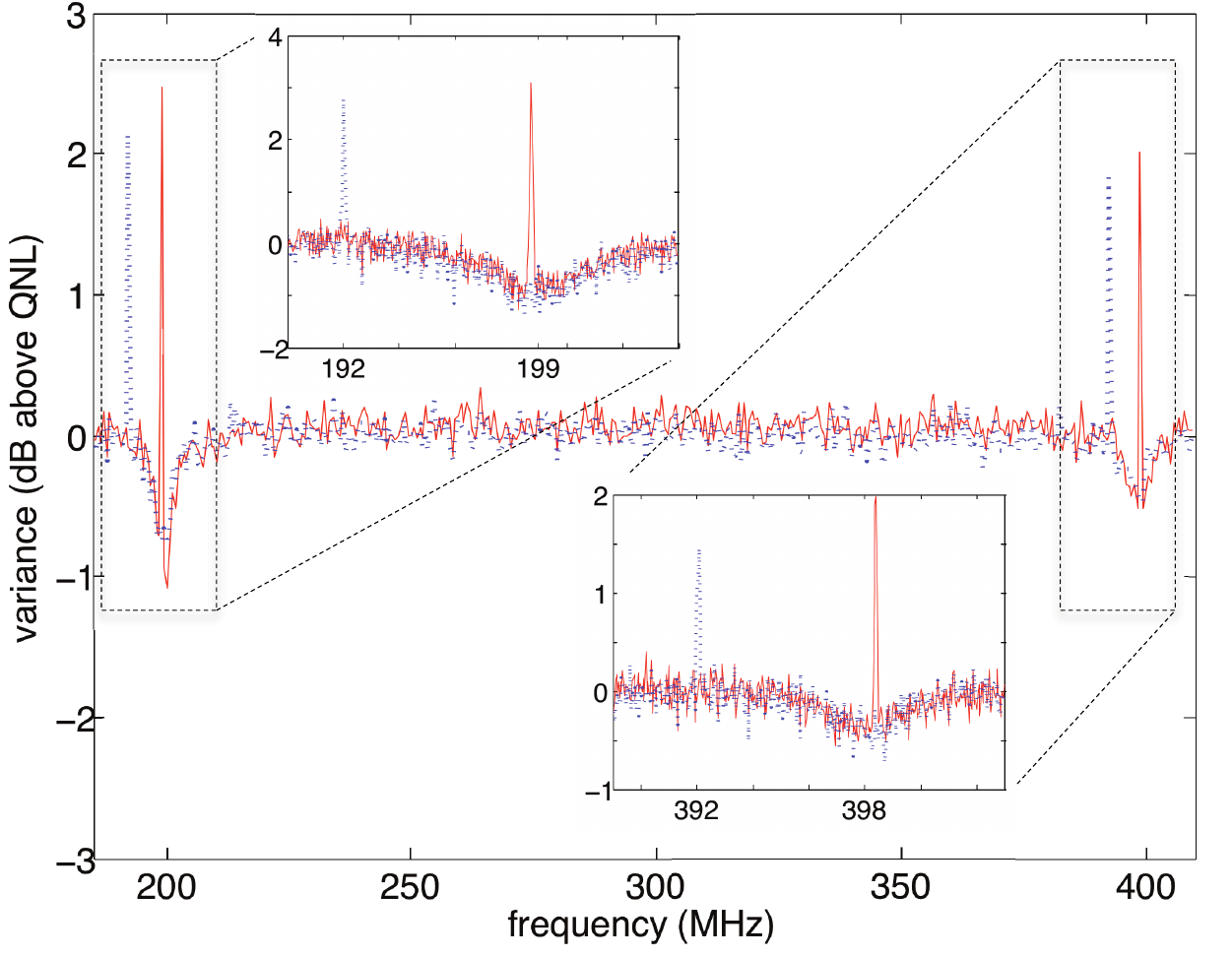}
   \caption{Measurements of FDM communications over the first and second resonances of the squeezing comb (color online).}
  \label{comms}
\end{figure}

We illustrate the principle of multiplexed communication on the quantum channel by implementing FDM on the first two resonances of the squeezing spectrum.  Two independent sinusoids are generated, combined electrically and used to phase modulate the OPO output as illustrated in Fig. \ref{schem}.  Subsequent detection and analysis completes the scheme.  Figure \ref{comms} shows the measured frequency spectrum of the scheme.  The solid (dashed) trace shows measurements relative to the QNL when the carrier frequencies are (not) aligned with the resonances in the squeezing spectrum, $199$ and $398$ MHz ($192$ and $392$ MHz) respectively.  The improved signal to noise ratio ($V_s/V_n\approx1.46$ vs $V_s/V_n\approx1$) , and hence channel capacity ($C\approx0.65$ vs $C\approx0.5$), when the signaling and squeezing spectra are aligned is clearly observed.  The insets to Fig. \ref{comms} show a separate zoomed-in view of each independent frequency band. 

\section{Conclusion}

In summary,  we have reported on the first time-resolved, simultaneous observation of the first dozen teeth in a 2.4~GHz comb of vacuum squeezing produced by a sub-threshold OPO, and demonstrated multiplexed communications on that channel.  We have shown theoretically that a quantum communications channel supported by such a comb of vacuum squeezing would have a greater channel capacity per photon than one supported by a source of broadband squeezing with the same analogue bandwidth. These arguments carry over directly to the multiplexed distribution of quantum entanglement. If we consider quantum communication protocols that include photon counting then our system has the added advantage of producing well resolved frequency modes that can be cleanly separated through optical means and distributed to different photon counting modules.

{\it Acknowledgments -} This work was supported by the Australian
Research Council.

\end{document}